\documentclass{ws-procs9x6}
\usepackage{amssymb}
\begin{document}

\title{Modular symmetry, twisted sectors and flavour}

\author{Thomas Dent}

\address{Michigan Center for Theoretical Physics, \\
University of Michigan, Ann Arbor MI 48109-1120\\ 
}

\maketitle

\abstracts{We investigate the implications for fermion mass models in
heterotic orbifolds of the modular symmetry mixing twisted states 
localized at different fixed points. We show that, unlike in the case 
of continuous gauge symmetries, the mass eigenstates do not mix under 
the symmetry; thus coupling constants in the low-energy theory are 
invariant functions of the moduli.}

\section*{Introduction}
Twisted sectors of heterotic orbifolds
have long been of 
phenomenological interest, since their massless states are automatically
localized at the fixed points, resulting in effective couplings 
exponentially suppressed by the distance between different 
points\cite{Ibanez:1986ka}. In the 
field theory limit of large compactification radii, such nonlocal couplings 
vanish, but they acquire nonzero values through nonperturbative effects on 
the worldsheet involving stretched strings.
Attempts have been made to explain the hierarchy of fermion masses 
(which generally requires some Yukawa couplings to be orders of magnitude 
smaller than unity) and the CKM charged current mixing of quarks, by using 
this suppression\cite{Casas:1992zt}.

The usual route to explaining small Yukawa couplings in a natural way is a 
(broken) flavour symmetry, so one might ask whether there is a symmetry
associated with this suppression of nonlocal interactions. In fact, the
target-space modular invariance of heterotic strings\cite{Dijkgraaf:jt}
might play such a role. This symmetry can be thought of as the extension 
of T-duality to a $d$-dimensional torus, where $d\geq 2$. In the simplest 
case $d=2$, 
the path integral for strings propagating on a background with modulus 
$T\sim V + iB$, where $V$ is the torus volume in string units and $B$ is 
the antisymmetric tensor field ($B_{MN}\equiv B\epsilon_{MN}$ where $M$, 
$N$ label the torus coordinates), is equivalent to that on a background
$T'$, where $T'=(\alpha T -i\beta)/(i\gamma T+\delta)$, $\alpha,\beta, 
\gamma, \delta \in \mathbb{Z}$, $\alpha\delta-\beta\gamma=1$. This 
(P)SL$(2,\mathbb{Z})$ symmetry is generated by $\mathcal{S}\colon 
T\mapsto 1/T$, $\mathcal{T}\colon T\mapsto T+i$. For 6-dimensional 
orbifolds, the full duality group may be larger but includes three 
PSL$(2,\mathbb{Z})$ subgroups acting on diagonal moduli $T^I$, 
$I=1,\ldots,3$. Correlation functions involving twisted states,
which are related to Yukawa couplings, are then invariant, provided 
that the states in a given sector have the nontrivial 
transformation\cite{Lauer:1990tm}
\begin{equation}
\sigma_i \mapsto U_{ij} \sigma_j,\ U^\dag U=1
\end{equation}
where $U$ depends on $\alpha$, $\beta$, $\gamma$, and $\delta$ 
but\footnote{Up to a overall complex phase within each twisted sector.} 
not on $T$. 
This discrete 
nonabelian symmetry relates states with the same quantum numbers, acting 
as a horizontal symmetry, which may be relevant for the flavour problem.

Previous investigations of mass textures used models with more than three 
matter generations massless at tree level, and exploited the freedom to 
pick a subset of the fixed point states to correspond to the MSSM matter 
fields, under the assumption that the remaining states became unobservably 
heavy. 
Physical observables in such a framework are, in general, not invariant 
functions of $T$ since MSSM states transform into heavy states under 
some group elements\cite{Lebedev:2001qg}. However, it can be shown that 
such behaviour is inconsistent with modular invariance of the mass term 
in the Lagrangian, since there must be off-diagonal masses in a basis 
where supposedly ``light'' and ``heavy'' states are mixed by the 
symmetry\cite{Dent:2001mn}. 
One can find a basis where some elements of SL$(2,\mathbb{Z})$ act 
diagonally, but other elements inevitably mix states within a twisted sector.
We find that (except in very special cases) the resulting mass eigenstates 
are SL$(2,\mathbb{Z})$ invariant, $T$-dependent linear combinations of the 
original fixed point states. Thus, even when the modular symmetry is
spontaneously broken, the couplings of the light states are invariant 
functions of moduli.

\section*{Invariance of mass eigenstates}
Our starting points are the existence of the unitary transformation
$U$ and the invariance of the part of the $d\!=\!4$ effective Lagrangian 
bilinear in matter fields $C_i$, which will result in mass terms: 
\begin{equation}
-\mathcal{L}_{\rm m} = \frac{1}{2} C_i C_j y_{ijkl}(T) H_k \prod_m 
X_m^{a_{lm}} + {\rm h.\,c.}
\equiv C_i C_j M_{ij}(T,H,X) + {\rm h.\,c}.
\end{equation}
where $i$ runs over complete twisted sectors\footnote{The $C_i$ are 
left-handed Weyl fermions, with $M_{ij}=M_{ji}$ a complex symmetric 
matrix: this notation allows for both Dirac and Majorana masses. 
Gauge indices are suppressed.}.
Here, $y_{ijkl}$ are $T$-dependent Yukawa couplings, and we include an 
arbitrary number of Higgses $H_k$ and SM singlets $X_m$ which may acquire 
v.e.v.'s in the stabilized vacuum; all fields may have nontrivial 
modular transformation properties. 

To determine whether it is possible to separate the states into two 
sets, light (observable) and heavy, which intermix under 
SL$(2,\mathbb{Z})_T$, we must find the modular transformation of the 
mass eigenstates, given the matrices $U$ acting in the fixed point basis. 
When general off-diagonal mass matrices are considered, such transformations 
quickly become unwieldy: so instead of the fixed point basis, we start 
with a basis where the modular transformation that we wish to consider 
acts as a diagonal matrix of phases: $\Gamma \in {\rm SL}(2,\mathbb{Z})
\colon\ C_i \mapsto e^{i\zeta_i} C_i$. 
This can always be done, because unitary matrices are normal, and does 
not introduce additional $T$-dependence.
Then, since $\mathcal{L}_{\rm m}$ necessarily transforms into itself, we 
have in this basis
\begin{equation}
M_{ij} \mapsto e^{-i(\zeta_i+\zeta_j)} M_{ij},\ V_{jp}\mapsto e^{i\zeta_j}
V_{jp}
\end{equation}
where $V$ is $T$-, $H_k$-, and $X_m$-dependent unitary matrix satisfying 
$V^TMV\equiv {\rm diag}(m_i(T,H,X))$\footnote{The invariant functions $m_i$ 
may be complex, depending on the structure of $M$.}. 
Then the mass eigenstates $\Psi_p= V^\dag_{pj}C_j$ are invariant when 
the $C_i$, $T$, $H_k$ and $X_m$ are all transformed under $\Gamma$. If
we rewrite the theory in the $\Psi$ basis, all coupling constants must be
invariant functions of $T$, $H_k$, and $X_m$, following from the invariance
of the 4d effective Lagrangian.

There might be exceptions to this result if the off-diagonal 
``light-heavy'' mass-terms vanish exactly, or when the modular 
transformation acts on the $C_i$ as a permutation. However, the first
possibility could only occur at isolated points in moduli space\footnote{As
discussed later, off-diagonal terms cannot in general be removed by 
redefinition.}, and the second is not realised in any known orbifold.

The case of antigenerations which acquire masses by pairing 
up with matter generations at energies somewhat below the string scale 
can be treated by the above method, but one can also think of it in two 
stages at widely-separated mass scales. First, consider a 
$N_g+M$-by-$M$-dimensional mass matrix, where $M$ is the number of
paired (anti)generations, with $M$ large mass eigenvalues. Then the 
diagonalising matrices are determined only up to an $N_g$-dimensional 
unitary redefinition of the remaining massless generations. The main 
result remains: if a modular symmetry mixed massless and heavy fields, 
then there would necessarily be $T$-dependent 
off-diagonal mass terms, which could not be set to zero except
at isolated values of $T$, hence we could not actually be in the basis of 
light and heavy fields.

\section*{Discussion}
Our result, which generalises that of\cite{Dent:2001mn}, 
may appear unexpected compared to the analogous situation for
continuous gauge symmetry, where for example in the case of SU$(2)_W$ the
mass eigenstates $t$ and $b$ (neglecting the CKM mixing) are mixed by
some group elements and SU$(2)$ appears to be explicitly broken below the 
top mass. In that case, starting from an arbitrary Higgs v.e.v.\ of magnitude 
$v$, we are free to change basis by a global SU$(2)$ redefinition to 
obtain a diagonal mass matrix ({\em i.e.} unitary gauge), and the group acts
in an identical way in the new basis. But for SL$(2,\mathbb{Z})_T$, the 
fixed point basis is special, and the group acts differently if one changes 
basis. One cannot in general redefine either $T$ or the $C_i$ by a group 
element to obtain a (block-)diagonal mass matrix: the diagonalisation 
requires $V$ to be explicitly modulus-dependent, which affects the 
transformation properties of the mass eigenstates.

The main lesson is that the freedom to assign MSSM fields to particular 
fixed points, and discard other states, is an illusion. For 
self-consistency, one should take the light fields to be invariant 
$T$-dependent combinations of twist states, as happens automatically in most 
cases. Modular invariance is thus a useful tool for checking the consistency
of toy model results. More realistic models with three light generations, in
which there is need to decouple extra states, most likely would 
require more complicated string constructions which would break part or all 
of the modular symmetry, uniquely determining the states that remain. One 
might argue that choosing a non-modular invariant set of states from a larger 
spectrum could mimic this situation, but the Yukawa couplings are unlikely
to take the same form in the more realistic case. 

%

\section*{Acknowledgments}
The author acknowledges useful commments by Andr{\' e} Lukas, and the 
hospitality of the Aspen Center for Physics where part of this work was 
done.



\end{document}